\documentclass[aps,pra,onecolumn,groupedaddress,showpacs]{revtex4}

\usepackage{amssymb,amsmath}
\usepackage{graphicx}

\newcommand{\om}{\omega}

\newcommand{\pa}{\partial}

\begin{document}

\title{ Two-component nonlinear waves}

\author{G. T. Adamashvili}
\affiliation{Technical University of Georgia, Kostava str.77, Tbilisi, 0179, Georgia.\\ email: $adamash77@gmail.com.$ }

\begin{abstract}

The generalized equation for the study of two-component nonlinear waves in different fields of physics is considered. In special cases, this equation is reduced to a set of the various well-known equations describing nonlinear solitary waves in the different areas of physics. Using both the slowly varying envelope approximation and the generalized perturbation reduction method, the generalized equation is transformed into the coupled nonlinear Schr\"odinger equations and the two-component nonlinear solitary wave solution is obtained. Explicit analytical expressions for the shape and parameters of two-component nonlinear wave consisting of two breathers oscillating with the sum and difference frequencies and wave numbers are presented. The solution of the generalized equation coincides with the vector $0\pi$ pulse of the self-induced transparency.

\vskip+0.2cm
\emph{Keywords:} Two-component nonlinear waves, Generalized perturbation reduction method, Generalized equation for nonlinear solitary waves.
\end{abstract}


\maketitle

\section{Introduction}

Physical phenomena can be conditionally divided into two main kinds. The first kind includes such physical phenomena that can take place only in any one (or a few) physical system, requiring specific conditions that can only be realized in that particular system. Examples of such effects are superfluidity and superconductivity, among others. The second kind includes physical phenomena that can take place in many completely different physical systems, regardless of their specific properties and physical nature. Such phenomena are of particular interest because they characterize the general fundamental properties of matter. The latter phenomena include the formation of nonlinear solitary waves, which occur in completely different areas of physics for physical quantities with different natures. Examples include the strength of the electric field of an optical wave, the deformation tensor of an acoustical pulse, the deviation of the water surface from the unperturbed state, etc. Nonlinear solitary waves are very interesting physical objects that describe highly excited nonequilibrium states of a system localized in space and time, in which the physical parameters characterizing the system remain unchanged. Unlike linear waves, solitary nonlinear waves can transfer wave energy over considerable distances without significant losses. Nonlinear solitary waves have long been a subject of study, and a significant number of works have been published; nevertheless, they are still relevant and their intensive research continues. These waves were first discovered on the surface of liquids, but were subsequently also investigated in other areas of physics, including optics, condensed matter, plasma, field theory, acoustics, metamaterials, quantum dots, graphene, etc. These nonlinear waves in different media and different physical conditions have identical properties, although they are solutions of completely different nonlinear partial differential equations, such as the Korteweg de Vries equation, the Boussinesq equation, the Benjamin-Bona-Mahony equation, the Sine-Gordon equation, the Bloch-Maxwell system of equations, the Hirota equation, the nonlinear Schr\"odinger equation and many other equations (see, for instance, Refs.[1-7]).

Nonlinear solitary waves of two main types are considered: resonant and non-resonant nonlinear waves. Among nonlinear solitary waves, one-component (scalar) and two-component (vector) waves are the most common. They differ significantly from each other in terms of their shapes, parameters and physical properties. Single-component waves in the process of propagation retain not only their energy but also their profile. One-component nonlinear solitary waves include a scalar soliton, a scalar breather, and other scalar nonlinear waves. Two-component solitary nonlinear waves are more complex formations and have an internal structure. Two-component waves include a vector soliton, a vector breather, or a breather molecule. A vector breather is a bound state of two scalar breathers that oscillate at different frequencies, have the same velocities, propagation directions, and polarizations, or have mutually perpendicular polarizations, for example, for waveguide modes. In the process of propagation, the breathers that make up the vector breather interact with each other and exchange energy, although they propagate as a whole.

For the analytical description of one-component and two-component nonlinear solitary waves, the use of various mathematical methods is required. This is because more functions and parameters are required to study two-component nonlinear waves than to study single-component nonlinear waves. For example, the perturbative reduction method [8, 9] is often used to study single-component nonlinear waves, utilizing one auxiliary complex function and two real parameters. Via this method, it is impossible to study two-component nonlinear waves. For two-component nonlinear waves, a generalized version of the perturbative reduction method  was developed [10-12]; this uses two complex auxiliary functions and eight parameters.

One-component and two-component nonlinear solitary waves can be formed in both different and identical physical systems. For the implementation of each, appropriate physical conditions are required. Some effects cannot be described using one-component nonlinear waves, and the concept of two-component waves becomes necessary. The most striking example of such effects is self-induced transparency (SIT) [13, 14]. After the discovery of this effect, for half a century it was believed that the main SIT pulses are a scalar soliton ($2\pi$ pulse) and a scalar breather (scalar $0\pi$ pulse). However, after the development of a generalized perturbative reduction method [10-12], enabling the study of two-component waves, it was discovered that the physical picture of the SIT is qualitatively different. In the theoretical description of SIT, the second derivative of the strength of the electric field of the pulse concerning the spatial coordinate and time in Maxwell's wave equation plays a fundamental role. However, in the traditional theory of SIT formulated by McCall and Hahn, the second derivatives were neglected, or were taken into account according to the perturbation theory only as small corrections of the first derivatives. Through the generalized perturbative reduction method it became possible to consistently take into account the role of second derivatives in Maxwell's wave equation. It was shown that the second derivatives describe the interaction between scalar breathers and the formation of their bound state. As a result, it was found that the SIT scalar breathers pair to form a vector breather and, therefore, the main SIT pulses are, along with the scalar soliton ($2\pi$ pulse), a vector breather (vector $0\pi$ pulse). The vector $0\pi$ pulse is a vector breather that consists of a bound state of two breathers of the same polarization, one of which oscillates with the sum, and the second with the difference, of frequencies and wave numbers [10-12, 15]. Subsequently, such a vector breather was studied in acoustic SIT, and then in many other areas of physics for quantities of a completely different nature, which are described by such nonlinear equations as, for instance, the Benjamin-Bona-Mahony equation, the Sine-Gordon equation, the system of Bloch-Maxwell equations, the Boussinesq equation or its various modifications, the nonlinear Schr\"odinger equation, the Born-Infeld equation, the Hirota equation, etc. To study vector breather in these equations, the generalized perturbative reduction method was also used [16-21].

In view of the fact that the same phenomenon, namely the formation of a vector breather consisting of two breathers oscillating with the sum and difference of frequencies and wave numbers, can be observed in completely different areas of physics, for quantities of completely different physical nature, which are described by different nonlinear equations, it follows that the existence of such vector breathers is a general fundamental characteristic of matter. Consequently, a natural question arises: is it possible to describe such vector breathers in completely different areas of physics using one generalized equation?

In the present paper, precisely such an equation is proposed that makes it possible to describe vector breathers from various fields of physics in a unified way. This equation is solved using the generalized perturbative reduction method.

The rest of this paper is organized as follows: Section 2 is devoted to the generalized equation for slowly varying complex envelope functions, and using the generalized perturbation reduction method, we will transform Eq. (1) to the coupled nonlinear Schr\"odinger equations for auxiliary functions. In Section 3, the explicit analytical expressions for the shape and parameters of two-component nonlinear pulse will be presented. Finally, in Section 4, we will discuss the obtained results.

\vskip+0.5cm

\section{The generalized perturbative reduction method}

The generalized equation containing the second and fourth-order derivatives in the space coordinates and time which describe pulses can take on the following form:
\begin{equation}\label{ge}
\alpha \frac{\partial^{2} U}{\partial t^{2}}+\beta\frac{\partial^{2} U}{\partial z^{2}}+ \gamma\frac{\partial^{2} U}{{\partial z}{\partial t}} +\delta\frac{\partial^{4} U}{\partial z^4} +\mu \frac{\partial^{4} U}{{\partial z^{3}}{\partial t}}+\nu \frac{\partial^{4} U}{{\partial z^{2}}{\partial t^{2}}}+\rho \frac{\partial^{4} U}{{\partial z}{\partial t^{3}}}+\vartheta \frac{\partial^{4} U}{\partial t^{4}}=F_{non},
\end{equation}
where the nonlinear term of Eq.(1) is given by
\begin{equation}\label{non}
F_{non}=- G   \frac{\partial^{2} U^{3}}{\partial z^{2}} -  A (\frac{\partial U}{\partial t})^{2}   \frac{\partial^{2} U}{\partial z^{2}} - \sigma (\frac{\partial U}{\partial z})^{2}   \frac{\partial^{2} U}{\partial t^{2}}     + B \frac{\partial U}{\partial t}\frac{\partial U}{\partial z} \frac{\partial^{2} U}{\partial t   \partial z}-\alpha_{0}^{2} \sin U +F  \frac{\partial^{2} P^{(2P)}}{\partial t^{2}},
\end{equation}
$U(z, t)$ is a real function of space coordinate $z$ and time $t$ and represents the wave profile, while  $\alpha,\;\beta,\;\gamma,\;\delta,\;\mu,\;\nu,\;\rho$, $\vartheta$, $G$, $A$, $\sigma$, $B$, $\alpha_{0}^{2}$ and $F$ are arbitrary constants.

For the study of vector breather solution of Eq.(1), the pulse duration $T$ is of great importance. For pulses whose width satisfies the condition $\omega T >> 1$, we can use the slowly varying envelope approximation, where $\omega$  is the carrier wave frequency. In this case, the function $U(z, t)$ in Eq.(1) we can represent in the following form [22-26] \begin{equation}\label{eq1}
U(z,t)=\sum_{l=\pm1}\hat{U}_{l}(z,t) Z_{l},\;\;\;\;\;\;\;\;\;\;Z_{l}=e^{{il(k z -\om t)}},
\end{equation}
where $Z_{l}= e^{il(kz -\om t)}$ is the fast oscillating function, $\hat{U}_{l}$ represents the slowly varying complex envelope functions, which satisfied inequalities, \begin{equation}\label{swa}\nonumber
 \left|\frac{\partial \hat{U}_{l}}{\partial t}\right|\ll\omega
|\hat{U}_{l}|,\;\;\;\left|\frac{\partial \hat{U}_{l}}{\partial z
}\right|\ll k|\hat{U}_{l}|.
\end{equation}
and $k$ is the wave number of the carrier wave. For the reality of $U$, we set: $ \hat{U}_{+1}= \hat{U}^{*}_{-1}$.

Substituting Eq.(3) into Eq.(1) we obtain connection between parameters $\omega$ and $k$ in the form
\begin{equation}\label{dis1}
- \alpha {\omega}^{2} - \beta k^{2}+  \gamma k \omega  + \delta k^{4}  - \mu   k^{3} \omega  +\nu  \omega^{2} k^{2}  - \rho k  {\omega}^{3}  + \vartheta  {\omega}^{4}=0
  \end{equation}
and the nonlinear equation for envelope function $\hat{U}_{l}$:
\begin{equation}\label{enveq}
\sum_{l=\pm1}Z_l [ il A_{1} \frac{\partial \hat{U}_{l}}{\partial t} + A_{6}\frac{\partial^{2} \hat{U}_{l}}{\partial t^2}-il A_{2} \frac{\pa \hat{U}_{l}}{\pa z}
+A_{3} \frac{\pa^{2} \hat{U}_{l}}{\pa z^2} +A_{5} \frac{\pa^{2} \hat{U}_{l}}{{\pa z}{\partial t}}
+il A_{4}  \frac{\partial^{3} \hat{U}_{l}}{\partial z^{3}} + \delta \frac{\partial^{4} \hat{U}_{l}}{\partial z^{4}} +i l A_{7} \frac{\pa^{3} \hat{U}_{l}}{{\pa z^2}{\partial t }}
$$$$
+ \mu \frac{\partial^{4} \hat{U}_{l}}{{\partial z^{3}}{\partial t}}    +  i l A_{9} \frac{\pa^{3} \hat{U}_{l}}{{\pa z}{\partial t^2}}
+\nu  \frac{\pa^{4} \hat{U}_{l}}{{\pa z^2}{\partial t^2}}  + i l A_{8}\frac{\partial^{3} \hat{U}_{l}}{\partial t^3}  + \rho \frac{\pa^{4} \hat{U}_{l}}{{\pa t^3}{\partial z}}
+\vartheta \frac{\partial^{4} \hat{U}_{l}}{\partial t^{4}} ]=F_{non},
\end{equation}
where
\begin{equation}\nonumber
A_{1} =-2 \alpha \omega + \gamma  k  - \mu   k^{3} +2 \nu  \omega k^2   - 3 \rho  k  {\omega}^{2}   +4 \vartheta  {\omega}^{3},
$$
$$
A_{2}= - (2 \beta  k   - \gamma  \omega  -4 \delta  k^{3} +3 \mu   k^{2} \omega   - 2 \nu     k \omega^2  +\rho  \omega^3 ),
$$$$
A_{3}=\beta -6 \delta  k^{2} + 3 \mu  k \omega    - \nu  \omega^2,
$$
$$
A_{4} =   4 \delta  k -\mu  \omega,
$$
$$
A_{5}= \gamma  -3 \mu  k^{2}   +  4 \nu   k \omega   - 3 \rho  \omega^2,
$$
$$
A_{6}=\alpha -k^2 \nu   + 3 \rho k  \omega   -6 \vartheta  {\omega}^{2},
$$
$$
A_{7}= 3 \mu k   -2 \nu  \omega,
$$
$$
A_{8}= \rho  k    - 4 \vartheta \omega,
$$
$$
A_{9}=   2 \nu   k  - 3 \rho  \omega.
\end{equation}

In order to consider the two-component vector breather solution of Eq.(1), we use the generalized perturbative reduction method
(see, for instance [10-12, 15]) which makes it possible to transform the nonlinear equation (5) for the functions $\hat{U}_{l}$ to the coupled nonlinear Schr\"odinger  equations for auxiliary functions $f_{l,n}^ {(\alpha)}$. As a result, we obtain a two-component nonlinear pulse oscillating with the  difference and sum of the frequencies and wave numbers.

In the frame of this method, the complex envelope function  $\hat{U}_{l}$ can be represented as
\begin{equation}\label{gprm}
\hat{U}_{l}(z,t)=\sum_{\alpha=1}^{\infty}\sum_{n=-\infty}^{+\infty}\varepsilon^\alpha
Y_{l,n} f_{l,n}^ {(\alpha)}(\zeta_{l,n},\tau),
\end{equation}
where $\varepsilon$ is a small parameter,
$$
Y_{l,n}=e^{in(Q_{l,n}z-\Omega_{l,n}
t)},\;\;\;\zeta_{l,n}=\varepsilon Q_{l,n}(z-v_{{g;}_{l,n}} t),
$$$$
\tau=\varepsilon^2 t,\;\;\;
v_{{g;}_{l,n}}=\frac{\partial \Omega_{l,n}}{\partial Q_{l,n}}.
$$

Such a representation allows us to separate from $\hat{U}_{l}(z,t)$ in the more slowly changing functions $f_{l,n}^ {(\alpha)}(\zeta_{l,n},\tau)$.

It is assumed that the quantities $\Omega_{l,n}$, $Q_{l,n}$ and $f_{l,n}^{(\alpha)}$ satisfy the inequalities for any $l$ and $n$:
\begin{equation}\label{rtyp}\nonumber\\
\omega\gg \Omega_{l,n},\;\;k\gg Q_{l,n},\;\;\;
\end{equation}
$$
\left|\frac{\partial
f_{l,n}^{(\alpha )}}{
\partial t}\right|\ll \Omega_{l,n} \left|f_{l,n}^{(\alpha)}\right|,\;\;\left|\frac{\partial
f_{l,n}^{(\alpha )}}{\partial \eta }\right|\ll Q_{l,n} \left|f_{l,n}^{(\alpha )}\right|.
$$

The parameters $\Omega_{l,n}$ and $Q_{l,n}$ are characterize more slowly oscillations in time and space coordinate in comparison with carrier wave frequency $\omega$ and wave number $k$.

By substituting Eq.(6) into Eq.(5) for the left-hand side of the generalized equation we obtain
\begin{equation}\label{eqz}
\sum_{l=\pm1}\sum_{\alpha=1}^{\infty}\sum_{n=\pm 1}\varepsilon^\alpha Z_{l} Y_{l,n}[W_{l,n}
+ \varepsilon i J_{l,n}\frac{\partial }{\partial \zeta_{l,n} } - \varepsilon^2 i l h_{l,n}  \frac{\partial }{\partial \tau}
-\varepsilon^{2} Q^{2}_{l,n} H_{l,n}\frac{\partial^{2} }{\partial \zeta_{l,n}^{2}}+O(\varepsilon^{3})]f_{l,n}^{(\alpha)}=F_{non},
\end{equation}
where
\begin{equation}\label{cof}
W_{l,n}= l A_{1} n\Omega -  A_{6} {\Omega}^{2}+ l A_{2} n Q -A_{3}  Q^{2} +A_{5}  Q \Omega+ l A_{4} n Q^{3} + \delta  Q^{4} - l A_{7} n Q^{2} \Omega  - \mu Q^{3} \Omega+ l A_{9} n {\Omega}^{2}Q
$$$$
+\nu   Q^{2} \Omega^{2} - l A_{8} n {\Omega}^{3} - \rho  Q  {\Omega}^{3}  +\vartheta  {\Omega}^{4},
$$$$
J_{l,n}= - l A_{1} Q v_g  +2 A_{6}   n \Omega Q v_g  - l A_{2}  Q + 2 A_{3}   n Q^{2} - A_{5}   n Q (Q v_g +\Omega)  -l A_{4}  3  Q^{3}  - 4 \delta  n Q^{4}  +  l A_{7}  Q^{2}(Q v_g +  2 \Omega)
$$$$
+ \mu n Q^{3} ( Q v_g +3 \Omega) -  l A_{9}    Q \Omega ({\Omega} + 2    Q v_g ) -2\nu    n Q^{2} \Omega( Q v_g + \Omega)
$$$$
+  l A_{8} 3 {\Omega}^{2}  Q v_g + n {\Omega}^{2} \rho   Q  (\Omega  + 3 Q v_g ) - 4 \vartheta  n {\Omega}^{3}  Q v_g,
$$$$
h_{l,n}= -A_{1}  +2 l A_{6}  n\Omega - A_{5}l  n Q  + A_{7}   Q^{2}  +l \mu n Q^{3}- 2   A_{9} Q \Omega   - 2 l \nu n Q^{2} \Omega  + A_{8}  3 {\Omega}^{2}   +3 l \rho  n Q  {\Omega}^{2}  - 4 l \vartheta   n {\Omega}^{3},
$$$$
H_{l,n}=- A_{6}  v_g^{2}- A_{3}    + A_{5} v_g +  3 l A_{4}  n Q    + 6 \delta Q^{2}  + l A_{7}   n ( 2 Q v_g  + \Omega)  -3\mu  Q ( Q v_g + \Omega)+  l A_{9} n   v_g ( 2  \Omega + Q v_g )
$$$$
+\nu  ( Q^{2} v^{2}_g    +  4 \Omega  Q v_g  +  \Omega^{2} ) - l A_{8} 3 n \Omega  {v_g}^{2}  - 3\rho    \Omega   v_g (\Omega+ Q  v_g)  + 6 \vartheta {\Omega}^{2}  v^{2}_g.
\end{equation}

For the sake of simplicity, we omit $l$ and $n$ indexes for the quantities $\Omega_{l,n}$, $Q_{l,n}$, ${v_{g;}}_{l,n}$ and $\zeta_{l,n}$ in Eq.(8) and furthermore where this will not be messy.

The nonlinear term $F_{non}$ on the right-hand side of the equation (7) is of order to $\varepsilon^{3}$.

In order to follow the standard procedure for the asymptotic methods, we equate the terms with the same powers of $\varepsilon$ to zero. From Eq.(7), we obtain a series of equations. In the first order of $\varepsilon$, we have that when $ f_{l,n}^{(1)}\neq0 $  the connection between the parameters $\Omega_{l,n}$ and $Q_{l,n}$ has the form
\begin{equation}\label{diss}
l A_{1} n\Omega -  A_{6} {\Omega}^{2}+ l A_{2} n Q -A_{3}  Q^{2} +A_{5}  Q \Omega+ l A_{4} n Q^{3} + \delta  Q^{4} - l A_{7} n Q^{2} \Omega  - \mu Q^{3} \Omega+ l A_{9} n {\Omega}^{2}Q
$$$$
+\nu   Q^{2} \Omega^{2} - l A_{8} n {\Omega}^{3} - \rho  Q  {\Omega}^{3}  +\vartheta  {\Omega}^{4}=0.
\end{equation}

From Eq.(9) we obtain:
\begin{equation}\label{vg}
v_{{g;}_{l,n}}=\frac{  ln A_{2}  - 2 A_{3}    Q + A_{5}    \Omega  + 3ln A_{4}    Q^{2} + 4 \delta   Q^{3}  - 2 ln A_{7}  Q \Omega - 3 \mu  Q^{2} \Omega +  ln A_{9}     {\Omega}^{2}  + 2\nu    Q \Omega^2   -  {\Omega}^{3} \rho    }{- ln A_{1}   +2 A_{6}   \Omega   - A_{5}   Q   +  ln A_{7}  Q^{2}   +  \mu  Q^{3}    - 2 l n A_{9}    Q \Omega      -2\nu    Q^{2} \Omega   +  l n A_{8} 3 {\Omega}^{2}   + 3 {\Omega}^{2} \rho   Q    - 4 \vartheta  {\Omega}^{3}  }.
\end{equation}

From Eqs.(7), (8) and (9), in  the second order of $\varepsilon$, we obtain the equation $J_{l,n}=0$ for any indexes $l$ and $n$.

In the third order of $\varepsilon$,  the nonlinear equation (7), is given by
\begin{equation}\label{l}
\sum_{l=\pm1}\sum_{n=\pm 1}\varepsilon^{3} Z_{l} Y_{l,n}[ - i l h_{l,n}  \frac{\partial }{\partial \tau}- Q_{l,n}^{2} H_{l,n}\frac{\partial^{2} }{\partial \zeta_{l,n}^{2}}]f_{l,n}^{(1)}=F_{non}.
\end{equation}

Next, we considered the nonlinear term $F_{non}$ of the nonlinear wave equation (11). Substituting Eqs.(6) and (3) into Eq.(2) for the nonlinear term proportional to $Z_{+1}$, we obtain
\begin{equation}\label{non2}
 Z_{+1}\varepsilon^{3}[ \tilde{q}_{+}  | f_{+1,+1}^ {(1)} |^{2}+\tilde{r}_{+} | f_{+1,-1}^ {(1)} |^{2}     \} Y_{+1,+1} f_{+1,+1}^ {(1)}
 +  \tilde{q}_{-} | f_{+1,-1}^ {(1)} |^{2} +\tilde{r}_{-} | f_{+1,+1}^ {(1)} |^{2}       \}Y_{+1,-1} f_{+1,-1}^ {(1)}]
\end{equation}
and plus terms proportional to $Z_{-1}$, where we use the notations
$$
\tilde{q}_{\pm} =\frac{\alpha_{0}^{2}}{2} + 3 G(k\pm Q_{\pm})^{2}+(A +\sigma -B )(\om \pm \Omega_{\pm} )^{2}  (k \pm Q_{\pm} )^{2} \pm  \frac{\kappa^{(2p)}}{ \Omega_{\pm}},
$$
$$
\tilde{r}_{\pm}=\alpha_{0}^{2}+ 6 G(k \pm Q_{\pm})^{2}+2[ A (\om \mp \Omega_{\mp} )^{2} (k \pm Q_{\pm} )^{2}
     +  \sigma  ( \omega \pm \Omega_{\pm}  )^{2}     (k \mp Q_{\mp}  )^{2}
$$$$
 -B (  \om + \Omega_{+}  )   (  \om -\Omega_{-} ) (  k+ Q_{+}  )  (   k- Q_{-} )] \pm \frac{4 \kappa^{(2p)}}{\Omega_{+}-\Omega_{-}},
$$
$$
\kappa^{(2p)}= \frac{\pi \omega^{2}n_{0} |d_{0}|^{2}}{\hbar},
$$
$$
 Q_{+}=Q_{+1,+1}= Q_{-1,-1},\;\;\;\;\;\;\;\;\;\;\;\;\;\;\;\;\;\;\;\;\;\;\;\;\;  Q_{-}=Q_{+1,-1}= Q_{-1,+1},
$$
$$
\Omega_{+}=\Omega_{+1,+1}= \Omega_{-1,-1},\;\;\;\;\;\;\;\;\;\;\;\;\;\;\;\;\;\;\;\;\;\;\;\;\;  \Omega_{-}=\Omega_{+1,-1}= \Omega_{-1,+1}.
$$
$\hbar $ is Planck's constant,  $n_{0}$ and  $d_{0}$ are the concentration and the matrix element of electric dipole
moment of the transition between the energy levels of two level impurity atoms, respectively.

From Eqs.(7) and (12), in the third order of $\varepsilon$, we obtain the system of nonlinear equations
\begin{equation}\label{2eq}
  i \frac{\partial f_{+1,+1}^{(1)}}{\partial \tau} + Q_{+}^{2} \frac{H_{+1,+1} }{h_{+1,+1}} \frac{\partial^2 f_{+1,+1}^{(1)}}{\partial \zeta_{+1,+1} ^2}+(\frac{\tilde{q}_{+}}{ h_{+1,+1}}   | f_{+1,+1}^ {(1)}|^{2} + \frac{\tilde{r}_{+}}{ h_{+1,+1}} | f_{+1,-1}^ {(1)}|^{2} ) f_{+1,+1}^ {(1)}=0,
$$$$
   i \frac{\partial f_{+1,-1 }^{(1)}}{\partial \tau} + Q_{-}^{2} \frac{H_{+1,-1} }{h_{+1,-1}} \frac{\partial^2 f_{+1,-1 }^{(1)}}{\partial \zeta_{+1,-1}^2}+(\frac{\tilde{q}_{-}}{h_{+1,-1}}  |f_{+1,-1}^ {(1)}|^{2} +\frac{\tilde{r}_{-}}{h_{+1,-1}} |f_{+1,+1} ^ {(1)}|^{2} )  f_{+1,-1}^ {(1)}=0.
 \end{equation}

\vskip+0.5cm
\section{The two-component vector breather }

After transformation back to the space coordinate $z$ and time $t$, from Eqs.(13) we obtain the coupled nonlinear Schr\"odinger equations for the auxiliary functions $\Lambda_{\pm}=\varepsilon  f_{+1,\pm1}^{(1)}$ in the following form
\begin{equation}\label{pp2}
i (\frac{\partial \Lambda_{\pm}}{\partial t}+v_{\pm} \frac{\partial  \Lambda_{\pm}} {\partial z}) + p_{\pm} \frac{\partial^{2} \Lambda_{\pm} }{\partial z^{2}}
+q_{\pm}|\Lambda_{\pm}|^{2}\Lambda_{\pm} +r_{\pm} |\Lambda_{\mp}|^{2} \Lambda_{\pm}=0,
\end{equation}
where
\begin{equation}\label{pp4}
p_{\pm}=\frac{H_{+1, \pm 1}}{h_{+1, \pm 1}},\;\;\;\;\;\;\;\;\;\;\;\;\;\;\;\;\;
q_{\pm}=\frac{ \tilde{q}_{\pm}}{h_{+1, \pm 1}},\;\;\;\;\;\;\;\;\;\;\;\;\;\;\;\;\;
r_{\pm}=\frac{ \tilde{r}_{\pm}}{h_{+1, \pm 1}},\;\;\;\;\;\;\;\;\;\;\;\;\;\;\;\;\;
v_{\pm }= v_{g;_{+1,\pm 1}}.
\end{equation}

The solution of Eq.(14) is given by [10, 11, 15]
\begin{equation}\label{ue1}
\Lambda_{\pm }=\frac{A_{\pm }}{b T}Sech(\frac{t-\frac{z}{V_{0}}}{T}) e^{i(k_{\pm } z - \omega_{\pm } t )},
\end{equation}
where $A_{\pm },\; k_{\pm }$ and $\omega_{\pm }$ are the real constants, $V_{0}$ is the velocity of the nonlinear wave. We assume that
$k_{\pm }<<Q_{\pm }$  and $\omega_{\pm }<<\Omega_{\pm }.$

By combining Eqs.(3), (6) and (16), we obtain the two-component vector breather solution of the nonlinear wave equation (1) in the following form:
\begin{equation}\label{vb}
U(z,t)=R \;Sech(\frac{t-\frac{z}{V_{0}}}{T})\{   \cos[(k+Q_{+}+k_{+})z
-(\omega +\Omega_{+}+\omega_{+}) t]
$$$$
+\sqrt{\frac{p_{-}q_{+}-2 p_{+}q_{-}}{p_{+}q_{-}- 2p_{-}q_{+}}}  \cos[(k-Q_{-}+k_{-})z -(\omega -\Omega_{-}+\omega_{-})t]\},
\end{equation}
where
$R$ is the amplitude of the pulse. The connection between the width $T$ and the velocity $V_{0}$  of the two-component nonlinear pulse is determined as:
\begin{equation}\label{rrw}
T^{-2}=V_{0}^{2}\frac{v_{+}k_{+}+k_{+}^{2}p_{+}-\omega_{+}}{p_{+}},
\end{equation}
where the connection between parameters of the nonlinear wave has the form:
\begin{equation}\label{ttw}
 \omega_{+}=\frac{p_{+}}{p_{-}}\omega_{-}+\frac{V^{2}_{0}(p_{-}^{2}-p_{+}^{2})+v_{-}^{2}p_{+}^{2}-v_{+}^{2}p_{-}^{2} }{4p_{+}p_{-}^{2}}.
\end{equation}

\vskip+0.5cm

\section{Conclusion }

In this paper, we studied the two-component vector breather of the nonlinear wave equation (1) under the condition of the slowly varying envelope approximation Eq.(3).  We considered a nonlinear pulse with the width $T>>\Omega_{\pm }^{-1}>>\omega^{-1}$.
Using the generalized perturbation reduction method Eq.(6), Eq.(7) was  transformed to the coupled nonlinear Schr\"odinger equations (14) for the auxiliary functions $\Lambda_{\pm 1}$.  As a result, the two-component nonlinear pulse oscillating with the sum $\omega +\Omega_{+}$ $(k+Q_{+})$ and difference $\omega -\Omega_{-}$ ($k-Q_{-}$)  of the frequencies (wave numbers) Eq.(17) was formed. The dispersion relation and the connection between parameters $\Omega_{\pm}$ and $Q_{\pm}$ are determined from Eqs.(4) and (9).
The parameters of the nonlinear pulse from Eqs. (8), (10), (15), (18) and (19) were also determined. This solution coincides with the vector $0\pi$ pulse of the self-induced transparency [10, 11, 15].

The nonlinear wave equation (1) is a generalized equation which united several special cases.

Indeed, when  the coefficients in Eq.(1) satisfied the conditions
\begin{equation}\label{c1}\nonumber
\alpha=1,\;\;\beta=-C,\;\; \gamma=\delta=\mu =\nu =\rho =\vartheta=G=A=\sigma=B=F=0,
\end{equation}
this equation is transformed into the Sine-Gordon equation
\begin{equation}\label{sg}
\frac{\partial^{2} U}{\partial t^{2}}-C \frac{\partial^{2} U}{\partial z^{2}} =-\alpha_{0}^{2} \sin U.
\end{equation}
Eq.(20) was solved by means of inverse scattering transform [2, 4], by which it is possible to
obtain the complete solution of the equation in the form of the nonlinear solitary waves.

When  the coefficients in Eq.(1)
\begin{equation}\label{cp}\nonumber
\alpha=1,\;\;\beta=-C,\;\; \gamma=\delta=\mu =\nu =\rho =\vartheta=G=F=\alpha_{0}^{2}=0,
\end{equation}
are fulfilled, Eq.(1) is reduced to the Born-Infeld equation
\begin{equation}\label{bi}
\frac{\partial^{2} U}{\partial t^{2}}-C \frac{\partial^{2} U}{\partial z^{2}} =-  A (\frac{\partial U}{\partial t})^{2}   \frac{\partial^{2} U}{\partial z^{2}} - \sigma
(\frac{\partial U}{\partial z})^{2}   \frac{\partial^{2} U}{\partial t^{2}}     + B \frac{\partial U}{\partial t}\frac{\partial U}{\partial z} \frac{\partial^{2} U}{\partial t   \partial z}.
\end{equation}

When  the coefficients in Eq.(1) have the form
\begin{equation}\label{cp}\nonumber
\rho =\vartheta=A=\sigma=B=F=\alpha_{0}^{2}=0,
\end{equation}
Eq.(1) is transformed to the nonlinear cubic Boussinesq-type equation which incorporate the different
versions of the Boussinesq-type equations considered in Ref.[21]
\begin{equation}\label{bu}
\alpha \frac{\partial^{2} U}{\partial t^{2}}+\beta\frac{\partial^{2} U}{\partial z^{2}}+ \gamma\frac{\partial^{2} U}{{\partial z}{\partial t}} +\delta\frac{\partial^{4} U}{\partial z^4} +\mu \frac{\partial^{4} U}{{\partial z^{3}}{\partial t}}+\nu \frac{\partial^{4} U}{{\partial z^{2}}{\partial t^{2}}}
=- G   \frac{\partial^{2} U^{3}}{\partial z^{2}}.
\end{equation}

When  the coefficients in Eq.(1) satisfied the conditions
\begin{equation}\label{c1}\nonumber
\alpha=1,\;\;\beta=-\frac{1}{c^2},\;\;F=-\frac{4\pi}{c^2},\;\; \gamma=\delta=\mu =\nu =\rho =\vartheta=G=A=\sigma=B=0,
\end{equation}

Eq.(1) is reduced to the Maxwell nonlinear wave equation for two-photon SIT
\begin{equation}\label{ge2}
\frac{\partial^{2} U}{\partial t^{2}}-\frac{1}{c^2} \frac{\partial^{2} U}{\partial z^{2}}
=-\frac{4\pi}{c^2}  \frac{\partial^{2} P^{(2P)}}{\partial t^{2}},
\end{equation}
where $P^{(2P)}$ is the  the resonance polarization of two-level optical impurity
atoms or semiconductor quantum dots under the condition of the two-photon resonance transitions, $c$ is the velocity of light in vacuum.

The Sine-Gordon equation (20) describes the geometry of surfaces with Gaussian curvature, the Josephson transition, SIT, dislocations in crystals, waves in ferromagnetic materials, the some properties of particles and the nonlinear phenomena in the different areas of research: nonlinear optics, quantum dots, two-dimensional materials, plasma, metamaterials, and others [1-4]. The Born-Infeld equation (21) is used in the analysis of  the nonlinear interaction in electrodynamics, in the theory of strings, in atomic physics and others [1, 27-31]. The Boussinesq-type equations Eq.(22) is used to describe of many various phenomena, for instance, it is a model of dispersive nonlinear solitary waves in a one dimensional lattice, in shallow water under gravity, in geotechnical engineering practice [32, 33], water-wave in weakly dispersive medium such as surface waves in shallow waters, an ion acoustic waves [34], wave propagation of elastic rods, vibrations in a nonlinear string and many others (see for details Ref.[21]). The Maxwell wave equation (23) describes nonlinear coherent interaction between an optical pulse and material in which resonance optical impurity atoms or semiconductor quantum dots have been embedded. In particular, under the condition of the two-photon resonance transitions of SIT.

To summarize, we see that the considered nonlinear equation (1) gives us the opportunity to consider vector breathers oscillating with the sum and difference of frequencies and wave numbers in various fields of physics to describe absolutely various physical phenomena.

\vskip+0.5cm

\end{document}